\documentstyle[prd,preprint,aps,tighten]{revtex}

\newcommand\beq{\begin{equation}}
\newcommand\eeq{\end{equation}}

           \def\l{\lambda}
\def\L{\Lambda}  \def\m{\mu}      \def\n{\nu}        
             
       \def\t{\tau}

\def\CD{{\cal D}}

\def\slash#1{\,/\kern-7pt#1}
\def\rd{\partial}

\def\darr#1{\raise1.5ex\hbox{$\leftrightarrow$}\mkern-16.5mu #1}

\def\rds{/\kern-6pt\rd}
\newcommand{\be}{\begin{equation}}
\newcommand{\bea}{\begin{eqnarray}}
\newcommand{\ee}{\end{equation}}
\newcommand{\eea}{\end{eqnarray}}
\newcommand{\ba}[1]{\left(\begin{array}{#1}}
\newcommand{\ea}{\end{array}\right)}

\newcommand{\id}{\kern0.2em\rule{0.1mm}{0.71em}
                 \kern0.12em\rule{0.1mm}{0.71em}
                 \kern-0.27em\rule[0.68em]{0.27em}{0.1mm}
                 \kern-0.30em\rule{0.44em}{0.1mm}\rule{0.1em}{-1mm}}

\def\prl#1#2#3{Phys. Rev. Lett. {\bf {#1}} ({#2}) {#3}}

\def\plb#1#2#3{Phys. Lett. {\bf B{#1}} ({#2}) {#3}}

\def\npb#1#2#3{Nucl. Phys. {\bf B{#1}} ({#2}) {#3}}

\preprint{\vbox{ \hbox{KIAS-P98005} \hbox{SOGANG-HEP 246/98}
\hbox{hep-th/9811005}}}
\begin{document}
\draft
\title{Statistical entropy and $AdS/CFT$ correspondence in BTZ black holes}
\author{Seungjoon Hyun $^{(a)}$\footnote{email: hyun@kias.re.kr},
        Won Tae Kim $^{(b)}$ \footnote{email: wtkim@ccs.sogang.ac.kr} and
        Julian Lee $^{(a)}$\footnote{email: jul@kias.re.kr} }
\address{ $^{(a)}$ School of Physics, KIAS, Seoul 130-012, Korea\\
          $^{(b)}$Department of Physics and Basic Science Research Institute,\\
         Sogang University, C.P.O. Box 1142, Seoul 100-611, Korea}
\date{\today}
\maketitle
\vskip 2.0cm
\begin{abstract}
We study $AdS/CFT$ correspondence in the case of AdS$_3$.  
We obtain the statistical entropy of the BTZ black hole 
in terms of the correct central
charge and the conformal dimensions for 
the states corresponding to the BTZ black hole. We point out the difference between our method and the old fashioned approaches based on 
$SL(2,R)$ Wess-Zumino-Witten model or Liouville theory. 
\end{abstract}
\newpage
 Recently there have been great interests in
three-dimensional locally anti-de
Sitter ($AdS_3$) black hole solutions of
Ban\~{a}dos-Teitelboim-Zanelli (BTZ)\cite{btz}. It is relevant for
the study of higher-dimensional black-hole solutions of string/$M$ theory
whose near horizon
geometries are those of BTZ solutions\cite{hyun}.  
As is
well-known, the asymptotic symmetry of $AdS_3$ is generated by two copies 
of Virasoro algebra\cite{bh}, which indicates the structure of conformal 
field theory (CFT) at the two-dimensional boundary at
infinity. The central charge of the algebra is given by 
\beq
c={3\ell \over 2G},
\label{cc}
\eeq
where $G$ is the three-dimensional Newton's
constant and $\L= -{1 \over \ell^2}$ is the cosmological constant. 
By using this observation and identifying the BTZ black hole with mass
$M$ and angular momentum $J$ as the boundary conformal field with 
dimension ($h$, $\bar h$),
\bea
h = {1 \over 2}(\ell M + J  ), 
\label{diml}
\eea
\bea
\bar h = {1 \over 2}(\ell M - J ), 
\label{dimr}
\eea
Strominger has obtained\cite{strominger} the microscopic
counting of Bekenstein-Hawking black hole entropy of BTZ solutions
(and higher dimensional solutions)\footnote{See also ref.\cite{ka}, \cite{bir}.} using Cardy's
formula\cite{cardy}: 
\be
S=2\pi\sqrt{c h \over 6} + 2\pi\sqrt{c \bar h \over 6}, \label{card}
\ee
However, as was pointed out in the literature\cite{carlip}, when the conformal dimension of the ground state is nonzero, $c$ in this formula should be replaced by 
\beq
c_{eff}= c-24 \Delta_0
\eeq
where $\Delta_0$ is the conformal dimension of the ground state. In particular, it well known that $1+1$ dimensional Liouville theory, which is induced at the boundary at  infinity by $2+1$ dimensional gravity, belongs to this class of theories.  This tells us that the Liouville theory does not have enough degrees of freedom to account for the states contributing to the black hole entropy. 

It was argued in ref.\cite{martinec} that the gravity theory is a kind of thermodynamic theory describing macroscopic physics, and does not contain the microstates corresponding to a given macro state. In the case of $2+1$ dimensions, the gravity theory, which can also be formulated as $SL(2,L) \times SL(2,R)$ Chern-Simons theory, has no degrees of freedom in the bulk, and therefore is equivalent to  the $SL(2,R)$ Wess-Zumino-Witten(WZW) model at the boundary, which becomes the Liouville theory after imposing appropriate boundary conditions\cite{chd}. Therefore from this viewpoint, it is not surprising  that we cannot obtain the correct entropy in the context of the Liouville theory, which amounts to counting the microscopic states. 

On the other hand, we have an interesting proposal  by Maldacena\cite{mald}  on the equivalence between conformal field theory(CFT) on $p$-dimensional spacetimes and the supergravities(string theories)
on $AdS_{p+1} \times \cal K$,where $\cal K$ is a compact manifold,\footnote{  The supergravity on $AdS_3$, whose bosonic sector is discussed in this paper, can be considered as coming from that on  $AdS_3 \times T^4$ (or $K3$)\cite{str}, in the limit where all the higher Kaluza-Klein modes are neglected. Since these matter fields act as sources to various currents of the boundary theory, we lose all the informations on the microscopic degrees of freedom except the stress-energy tensor by considering the $AdS_3$ gravity. Hence the arguments in ref.\cite{martinec} presented above. } which also appears in the context of M(atrix)  theory\cite{hyun1}. Gubser, Klebanov, and Polyakov in Ref.\cite{gkp} and Witten in Ref. \cite{witten} elaborate this correspondence further by identifying the partition function of the supergravity subjected to appropriate boundary conditions with the generating functional of connected Green's functions of the corresponding conformal field theory:
\beq
\int \CD\phi \exp(iI[\phi])=\left\langle\exp i\int_{\partial M} \phi_0 O \right\rangle_{CFT}  \label{dual}
\eeq   
where $I(\phi)$ is the classical action of the bulk theory, $\phi_0$ is  the boundary value of the bulk field $\phi$, $O$ is an operator in the boundary conformal field theory, and the path integral on the left hand-side is  only over the fields whose boundary value is given by $\phi_0$. That is, the boundary value of the bulk field is {\it fixed}, and acts as a source for the generating functional of the dual conformal field theory. On the other hand, the boundary  Liouville description in the case of the gravity in 2+1 dimensions is just a change of variable, so this correspondence can be written as:
\beq
\int \CD\phi \exp \int i ( j\phi_0+ I[\phi] )=\int \CD\phi_0 \exp \int_{\partial M}i ( j  \phi_0 + S_b[\phi_0])   \label{liou}
\eeq  
where $S_b[\phi_0]$ is the classical action for the boundary  Liouville theory.
Here we integrate over all possible boundary values $\phi_0$ since they are the dynamical degrees of freedom of the theory.  The difference of (\ref{dual}) from (\ref{liou}) becomes most drastic when one realizes that the quantum correction on the left-hand side of (\ref{dual}) is $1/N$ correction from the viewpoint of the dual conformal field theory on the right-hand side, where $N$ is the number of field content, such as size of matrices in the case of super Yang-Mills.\footnote{In the string theory picture, $N$ is just the number of branes.}  In particular, in the large $N$ limit, the left hand-side of (\ref{dual}) can be replaced by the value of the integrand at the classical solution with the given boundary condition,
\beq
\exp(iI[\phi_{cl}])=\left\langle\exp i\int_{\partial M} \phi_0 O \right\rangle_{CFT} \label{ln}
\eeq  
 That is, the classical action on the gravity side is the {\it quantum} effective action for the dual conformal field theory! On the other hand, the classical action for the gravity maps into classical action for the Liouville theory via $SL(2,R)$ WZW model.  

 In spite of these facts, there have been  widespread misconceptions that the Maldacena  type of duality is equivalent to the boundary $SL(2,R)$ WZW or Liouville theory description of the gravity in the case of 2+1 dimensions. We think this error is responsible for the most part of the confusion on the issue of the statistical entropy of black hole in 2+1 dimensions, such as  unitarity problem,  the effective central charge being too small, so on.   In particular, we see that in the light of the correspondence (\ref{dual})), (or (\ref{ln}) in the case of large $N$), it is quite misleading to consider the quantum Liouville theory in discussing the entropy of the BTZ black hole. The boundary action is already the quantum generating function of the dual conformal field theory, so why quantize it again?    

In this paper, we study $AdS/CFT$ duality  (\ref{ln}) in the case of $AdS_3$. We calculate the central charge, and also the conformal dimension of the BTZ black hole. One can then  read the statistical entropy of BTZ black hole, by adopting the approaches given in \cite{gkp}\footnote{The central charge has been obtained in \cite{henningson} by adopting the approach suggested in \cite{witten} which is closely related to ours.}
  This, in turn, supports the equivalence between conformal field theories on $p$-dimensional spacetimes and the supergravities(string/M theories) on $AdS_{p+1}$ advocated in \cite{mald}.

Now, consider the relation (\ref{ln}) for the case of 2+1 dimensional gravity. Taking the coordinates $(z, x^0, x^1)$ where $z$ direction is normal  and $x^a (a=0, 1)$ directions are parallel to the boundary surface, we have:  
\beq 
\exp(iI(\bar h_{a b}))=Z[ h_{a b}]\equiv
\left\langle\exp {i \over 4 \pi}\int_{\partial M}  
h_{a b} T^{a b} \right\rangle_{CFT}.  \label{sou}
\eeq
Here $\bar{ h}_{a b}$ $(a,b=1,2)$ are the boundary values of the bulk metric components which are parallel to the boundary, whereas $h_{a b}$ are the sources for the stress-energy tensor of the dual conformal field theory.\footnote{The relation between the boundary metric and the stress-energy tensor of the conformal field theory was also discussed in ref.\cite{sn}.} They are related by Weyl rescaling, which will be discussed later. The boundary values of the other metric components, which will act as sources for vectors and scalars in the dual conformal field theory, are not relevant for our discussions and will be set to zero for simplicity.   
 The expected form 
of the quantum effective action  $S_{eff}[h_{a b}]$ on the
boundary is \cite{polyakov}:
\beq
\exp (-S_{eff}[h_{a b}]) \equiv Z[h_{a b}]= \exp\left[-{c \over 96 \pi} \int d^2 x  
R {1 \over \Box} R \right] \label{anomaly}
\eeq
when the path integral measure is regularized in a generally covariant way.\footnote{ This expression is in Euclidean form, whereas we will do computations with Lorentzian signature.  Inverse derivatives are to be understood as Green's functions with standard boundary conditions.}  
To the quadratic order in $h_{ab}$, this is equal to
 \bea
S_{eff} &=& {c \over 24 \pi} \int d^2 x 
[h_{--} \rd_-^{-1} \partial_+^3   
h_{--} + h_{++} \rd_+^{-1} \rd_-^3  h_{++}+4 h_{+-} (\partial_+ \partial_- h_{+-} -\partial_+^2 h_{--} - \partial_-^2 h_{++} ) \nonumber \\
&& \qquad \qquad + 2 h_{++} \partial_- \partial_+ h_{--}] 
\label{nono}
\eea
  
Let us now study the $AdS/CFT$ correspondence in the case of $AdS_3$. The 2+1 dimensional gravity action with the negative cosmological constant is given by
\beq
I = {1 \over 16 \pi G} \int d^3 x \sqrt{-g} (R + {2 \over \ell^2} ) 
+ {1 \over 8\pi G}\int_{\partial M} d^2 x \sqrt{ -g^{(2)}} K 
\label{act}
\eeq
where the last term is the extrinsic curvature term:
\beq
\int_{\partial M} d^2 x \sqrt{- g^{(2)}}K \equiv \int_{\partial M} 
d^2 x
\sqrt{-g^{(2)}}g^{(2)}_{\mu \n} \nabla^\m n^\n ,
\eeq
where $n^\m$ is the unit vector normal to the boundary, $\nabla_\m$ is 
the covariant derivative with respect to the background metric, and $g^{(2)}$ 
is the projection operator to the two dimensional boundary,
\beq
g^{(2)}_{\m \n} \equiv g_{\m \n} - n_\m n_\n. 
\eeq   
We now consider the $M=0,J=0$ black hole solution:
\beq
ds^2 = \bar g_{\m \n} dx^\m dx^\n = {\ell^2 \over z^2}(dz^2 - dx^+ dx^-) 
\eeq 
and expand the metric around this background:
\be 
g_{\m\n}=\bar g_{\m \n} + \bar h_{\m \n}={\ell^2 \over z^2} (\eta_{\m\n}+h_{\m\n}). \label{expan}
\ee
where $x^\pm\equiv t/\ell \pm \phi$,
with $\phi$ and $\phi+ 2\pi$ identified, is a
two dimensional coordinate parametrizing 
the cylindrical boundary.\footnote{It
is crucial that we identify $\phi$ periodically. Otherwise, this metric
describes $8GM=-1$ $AdS_3$ background in the so called Poincare 
coordinate system. In these coordinates, $z=0$ is only a portion of the boundary. Similar argument applies to  general BTZ solutions
discussed later. Unless we identify $\phi$ periodically, they all describe portions of $AdS_3$ spacetime.}
This coordinate system is the one we already mentioned above, where $z$ direction is normal and $t, \phi$ directions are parallel to the boundary surface.
  
After we substitute (\ref{expan}) into the action (\ref{act}), we get a divergent
constant, which is irrelevant and thus 
dropped, plus surface terms. As mentioned earlier, since we are interested only in $ h_{\pm \pm},
h_{+-}$, we take the boundary condition such that $h_{z\mu}=0$ at $z=0$, although these $h_{ab}$ propagate to the bulk and get mixed with other components of 
$h_{\m\n}$. 
After some lengthy calculations, the action 
(\ref{act}) becomes 
\beq
I= {1 \over 16 \pi G \ell}\int d^2 x  ({4 \over z^2} h_{+-}  +{4 \over z^2}  h_{++} h_{--} - {1 \over z} 
h_{++} \partial_z   h_{--} - {1 \over z}  h_{--} \partial_z  h_{++}+{2 \over  z} h_{+-} \partial_z h_{+-}) .  \label{alfin}
\eeq
 Now one has to equate this expression with the effective action of the
 boundary conformal field theory. Since we have 
\be
g_{\mu \n}={\ell^2 \over z^2}(\eta_{\m \n}+h_{\m \n}) \rightarrow g_{\mu \n}=\eta_{\m \n}+h_{\m \n},
\ee
after the appropriate Weyl rescaling, it is natural to identify $h_{ab}$, the two dimensional components of $h_{\m \n}$, as the source for $T_{a b}$ in the dual conformal field theory, as in (\ref{sou})  
  
 Since $z$ direction is normal to the boundary 
surface ,  we must remove  the $z$
derivative in (\ref{alfin})  using the equations of motion, whose detailed form is given in the Appendix A, in order to relate it with a quantity in the two dimensional conformal field theory.   Then we get 
\bea
I &=&  -{\ell\over 16\pi G}
\int d^2 x [  h_{--} \rd_-^{-1} \rd_+^3 h_{--} 
+ h_{++}\rd_+^{-1} \rd_-^3  h_{++} +4 h_{+-} (\partial_+ \partial_- h_{+-} -\partial_+^2 h_{--} - \partial_-^2 h_{++} ) \nonumber \\
&& \qquad \qquad + 2 h_{++} \partial_- \partial_+ h_{--} - {4 \over z^2} h_{+-} - {4 \over z^4} h_{++} h_{--} ]. \label{nonlo}
\eea
Note that the  last two terms are local expressions which  diverge as $z \to 0$. As
discussed in Ref.\cite{polyakov}, 
the local terms depend on the regularization of
the quantum measure and can be removed by adding local counter-terms. These local counter-terms in conformal field theory corresponds in gravity theory to boundary terms  which depend only on the intrinsic geometry of the boundary\cite{af}.   Indeed we see that the divergent local terms are proportional to the area of the boundary and therefore  removed by adding
\beq
I_{add} = {1 \over 8 \pi G} \int_{\partial M} d^2 x \sqrt{- g^{(2)}} \label{add}
\eeq 
  to the original action. Then the resulting expression, which is given by the rest of the terms in (\ref{nonlo}),  agrees exactly with (\ref{nono}) with the central charge given in (\ref{cc}).  In summary, we have shown that
\bea
I &=& {1 \over 16 \pi G} \int d^3 x \sqrt{-g} (R + {2 \over \ell^2} ) 
+ {1 \over 8\pi G}\int_{\partial M} d^2 x \sqrt{ -g^{(2)}} K + {1 \over 8 \pi G}\int_{\partial M} d^2 x \sqrt{- g^{(2)}} \nonumber \\
&=& {1 \over 24 \pi}{3 \ell \over 2G }  \int d^2 x  
R {1 \over \Box} R  \label{final}
\eea
to the quadratic order in $h_{ab}$.
 
Next we obtain the conformal dimensions of the states corresponding to the BTZ
black hole with general $M,J$. 
In a conformal field theory, the stress-energy tensor on a cylinder is expanded as
\bea
T_{++}(x^+) &=& \sum_n L_n \exp(-i n x^+) \cr
T_{--}(x^-) &=& \sum_n \bar L_n \exp(-i n x^-)
\eea
and the conformal dimensions $h, \bar h$ of a state $| h, \bar h \rangle $  is defined to be the eigenvalues of $L_0, \bar L_0$, so we have 
\beq
\langle h | T_{++}(x^+) | h  \rangle  = \langle h | \sum_n L_n \exp(-i n x^+)
 | h, \rangle 
=  \langle h | L_0  | h \rangle =  h   \label{condim}
\eeq
where we suppressed the label $\bar h$. A similar relation holds for $T_{--}$ and $\bar h$.
    The expectation 
values of $L_n$ for $n \neq 0$ vanish since $| h \rangle$ and $L_n |h \rangle$ have different
conformal dimensions and thus orthogonal. We can calculate the left-hand side of the equation (\ref{condim})
in the gravity side, and match with the right-hand side to obtain the conformal
dimension.  

In the coordinate where $h_{z\mu} \to O(z^2)$ as $z\to 0$, the expressions for $h_{ab}$ are given by 
\bea
h^{(\rm BTZ)}_{zz}=O(z^4), \cr
h^{(\rm BTZ)}_ {+-} = O(z^4), \cr
h^{(\rm BTZ)}_{\pm \pm} = {2G \over \ell}z^2 (M \ell \pm J)  
\eea
as discussed in the Appendix B.  Using these expressions, we now have
\bea
{ \sum_i \langle h, i | T_{++} | h, i \rangle \over \sum_i \langle h, i | h, i
\rangle }
&=& {4\pi \over \sqrt{-\det(\eta+h)}}{\delta S_{eff} \over \delta h^{++} }|_{h=h(\rm BTZ)} \cr
& =& 4 \pi \left(1+O(z^2)\right){\ell \over 8\pi G}\eta_{+-}^2 \left[ \partial_-^{-1}\partial_+^3 h_{--} + \partial_+ \partial_- h_{++} -2 \partial_+^2 h_{+-} \right]_{h(BTZ)} \cr 
&=& {\ell \over 8G} \left[z\partial_z (z^{-2}h_{--})+2\partial_+^2 h_{+-} -\partial_+ \partial_- h_{++} + 2 z^{-2} h_{++} \right]_{h(BTZ)} \cr \cr
&=& {1 \over 2}(M\ell+ J ) = {c \over 24} (8G)(M+ {J\over \ell}) \label{look}
\eea
where $i$ denotes quantum numbers which distinguish various states
contributing to the black hole entropy.  We used the equation of
motion in Appendix A 
in going back from the non-local expression in the second line to the
local form in the third line of (\ref{look}).  One can repeat exactly
the same kind of analysis for $T_{--}$. Thus we get the conformal dimensions
of the black hole as in (\ref{diml}) and (\ref{dimr}). 

We have shown explicitly that 
conformal dimensions ($h$, $\bar h$) of the BTZ black hole as  quantum
states are given by 
Eqs. (\ref{diml}) and (\ref{dimr}) 
with the central charge
(\ref{cc}). From this, one can easily read off that $AdS_3$ spacetime, which is
$8GM =-1$ and $8GJ=0$ case, corresponds to the quantum state with the conformal
dimensions (-${c\over 24}$, -${c\over 24}$). In the supersymmetric version of the
theory this corresponds to the Neveu-Schwarz ground state, which has anti-periodic
boundary condition\cite{henneaux}. On the other hand, the  
extremal black hole
with $M=J=0$ corresponds to the Ramond ground state with periodic boundary
condition\cite{henneaux}.  
This not only explains the Bekenstein-Hawking entropy of the three-dimensional BTZ 
black hole, but also explains the entropy of all the higher dimensional
asymptotically flat black hole solutions whose near horizon
geometries are those of BTZ solutions\cite{hyun,strominger}. 
 From higher dimensional point of view,  the
Bekenstein-Hawking entropy may be thought of 
as the degeneracy of states on near
horizon surface \cite{strominger} of higher dimensional black holes. 
 Since we believe that the unitarity of the boundary conformal field theory would be guaranteed by the 
underlying string theory with R-R background,  we can use the Cardy's
formula \cite{cardy} for the asymptotic growth of the number of the states,
\be
S=2\pi\sqrt{c h \over 6} + 2\pi\sqrt{c \bar h \over 6},
\ee
to get the entropy of BTZ black holes
\be
S=\pi\sqrt{\ell(\ell M+J) \over 2G} + \pi\sqrt{\ell(\ell M-J) \over 2G} ,
\ee
which agrees exactly with the Bekenstein-Hawking entropy.

\vskip 12pt

\noindent
{\bf Acknowledgement}

S.H. thanks the organizers and participants of the Amsterdam Summer
Workshop, `String Theory and Black Holes', and in particular, H. Verlinde.
 J.L. thanks Antal Jevicki for useful discussions. 
W.T.K. was supported by Korea Research Foundation.

 \appendix
\section{some relevant formulae}
In this appendix we display some relevant formulas for references.  
The bulk action for the Einstein gravity, the first term in (\ref{act}), becomes
a divergent constant plus a surface term after we substitute $g_{\m \n} = \bar
g_{\m \n} + \bar h_{\m \n}$ where $\bar g_{\m \n}$ is a solution to the classical
equation of motion. This surface term has the form:
\bea
16\pi G I_{\rm bulk} &=& \int_{\partial M} d^2 x\sqrt{- \bar g} \nabla^\m 
[-\nabla_\m\bar h + \nabla_\n\bar h_\m^\n  \nonumber \\
&+& {3 \over 4} \bar h_\m^\t\nabla \bar h - 
{1 \over 2} \bar h^{\n \l} \nabla \bar h_{\l \m} 
-  \bar h_{\m \n} \nabla_\l \bar h^{\n \l} 
- {1 \over 4} \bar h \nabla_m \bar h 
+ {3 \over 4} \bar h_{\n \l} \nabla \bar h^{\l \n} + 
{1 \over 4} \bar h \nabla_\n \bar h_\m^\n ] , \label{bb}
\eea
where $\bar h= \bar g^{\m\n} \bar h_{\m\n}$, and $\nabla$ is the background covariant derivative.  $\bar h_{\mu \nu}$ is related to $h_{\m \n}$ used in the text via Weyl rescaling 
\beq
\bar h_{\m \n} = {\ell^2 \over z^2}h_{\m \n}
\eeq
as can be seen from (\ref{expan}).  This term, together with the extrinsic curvature term, gives the  expression (\ref{alfin}). 
    
The bulk equations of motion, written in terms of $h_{\m \n}$, are given by
\bea
-\partial_\pm^2 h_{zz} + 2 \partial_\pm \partial_z h_{\pm z}  
-{2 \over z}\partial_\pm h_{\pm z} - \partial_z^2 h_{\pm\pm}+{1 \over z} \partial_z h_{\pm\pm} &=& 0, \cr  \cr
{2 \over z^2} \partial_z^2 h_{+-}-{2 \over z^3} \partial_z h_{+-} + 
{2 \over z^2} \partial_+ \partial_- h_{zz} -{1 \over z^3} \partial_z h_{zz} +{2 \over z^4}  h_{zz} &&\cr 
+{2 \over z^3}( \partial_-  h_{z +} + \partial_+ h_{z-}) 
-{2 \over z^2}( \partial_- \partial_z h_{z +}  +\partial_+ \partial_z h_{z-}) &=& 0, \cr \cr
-2\partial_\pm^2 h_{z\mp} + 2 \partial_+ \partial_- h_{z\pm} + 
2 \partial_\pm\partial_z h_{+-}- 2 \partial_\mp \partial_z h_{\pm\pm} 
- {1 \over z}\partial_\pm h_{zz} &=&0, \cr  \cr
4 \partial_+\partial_- h_{+-} - {2 \over z} (\partial_+ h_{-z}+\partial_- h_{+z}) 
+{2 \over z} \partial_z h_{+-}  -2 \partial_+ ^2 h_{--} - 
2 \partial_-^2 h_{++} - {h_{zz} \over z^2}&=& 0. \label{eom}
\eea
Using the last two equations of (\ref{eom}), we can remove 
$z$ derivatives in favor of derivatives in two dimensions to get the non-local expression (\ref{nonlo}).
  
 \section{BTZ Black Hole}
 The metric for BTZ black hole can be written as
\bea
ds^2 &=& -{l^2 \over z^2}(1-4G M z^2) d x^+ dx^- +2G \ell(M\ell+J)
(dx^+)^2 
+ 2G \ell(M\ell-J)  (dx^-)^2 \cr
&&+{\ell^2  \over z^2 (1 - 8 G M z^2 + ({4G J\over \ell})^2 
z^4)}dz^2,
\eea
where periodic identification of the angle coordinate $\phi$ is to be
understood. At the asymptotic infinity $ z \to 0$, one can consider this as a
small perturbation near $M=0$ black hole background:
\beq
\bar h_{zz} = 8G \ell^2 M + O(M^2 z^2, J^2 z^2) , \quad \bar h_{+-} = 2 G M l^2 , 
\quad \bar h_{\pm \pm} = 2G\ell(M\ell \pm J) . 
\eeq    
In order to use the formalism developed in this paper, we must go to the
coordinate where $h_{z\mu}$ vanishes at the asymptotic infinity $z \to 0$. For
our purposes it is enough to consider the leading order in $(M z)^2, (J z)^2$. 

 The linearized diffeomorphism is given by\cite{wald}
\beq
\delta \bar h_{\mu \nu} = 2 \bar \nabla_{(\mu} \xi_{\n)} =  \partial_\m \xi^\l
\bar g_{\n \l} +\partial_\n \xi^\l \bar g_{\m \l} +\xi^\l \partial_\l 
\bar g_{\m\n},  \label{lntr}
\eeq
where $\xi^\l$ is the infinitesimal transformation parameter and $\bar
\nabla_\l$ is the covariant derivative with respect to the background metric
$\bar g_{\mu \n}$. We easily see that in order to remove $\bar h_{z \mu}$, it is
enough to use 
\beq
\xi^z = -2 G M  z^3 + O(z^5), \label{trv}
\eeq
with other components vanishing. Then after performing the transformation (\ref{lntr}) with the parameter (\ref{trv}), we get
\beq
\bar h_{zz} = O(z^2) , \quad \bar h_{+-} = O(z^2), 
\quad \bar h_{\pm \pm} = 2G\ell(M\ell\pm J) . 
\eeq
We see that $\bar h_{\pm \pm}$ are invariant under this transformation. After Weyl rescaling
\beq
 h_{\m \n} = {z^2 \over \ell^2} \bar h_{\m \n} 
\eeq
we get
 \beq
 h_{zz} = O(z^4) , \quad  h_{+-} = O(z^4), 
\quad  h_{\pm \pm} = 2G{z^2 \over \ell} (M\ell\pm J) .  
\eeq  
We used these expressions to obtain the conformal dimension of the BTZ black
hole states in (\ref{look}). 



\newpage

\begin{thebibliography}{999999}
\bibitem{btz} M.\ Ban\~{a}dos, C.\ Teitelboim, and J.\ Zanelli,
\prl{69}{1992}{1849}.
\bibitem{hyun} S. Hyun, hep-th/9704005.
\bibitem{bh} J. D. Brown and M. Henneaux, Comm. Math. Phys. 104 (1986) 207.
\bibitem{strominger} A. Strominger,  hep-th/9712251, J. High Energy Phys. 02 (1998) 009.
\bibitem{ka} N. Kaloper, hep-th/9804062, \plb{434}{1998}{285}.
\bibitem{bir} D. Birmingham, I. Sachs and S. Sen hep-th/9801019, \plb{424}{1998}{275}.
\bibitem{cardy} J. A. Cardy, \npb{270}{1986}{186}.
\bibitem{carlip} S. Carlip, hep-th/9806026, Class. Quant. Grav. 15 (1998) 3609.
\bibitem{martinec} E. J. Martinec, hep-th/9809021.
\bibitem{chd} O. Coussaert, M. Henneaux, and P. van Driel, gr-qc/9506019, Class. Quant. Grav. 12. 
\bibitem{str} J. Maldacena and A. Strominger, hep-th/9804085, J. High Energy Phys. 12 (1998) 005 ; \break  E. J. Martinec, hep-th/9804111; J. de Boer, hep-th/9806104;  A. Giveon, D. Kutasov, \break and N. Seiberg,  hep-th/9806194,  Adv. Theor. Math. Phys. 2 (1998) 733.
\bibitem{mald} J.M. Maldacena,  hep-th/9711200, Adv. Theor. Math. Phys. 2 (1998) 231.
\bibitem{hyun1} S. Hyun, hep-th/9802026, \plb{441}{1998}{116}.
\bibitem{gkp} S. S. Gubser, I. R. Klebanov and A. M. Polyakov, 
 hep-th/9802109, \plb{428}{1998}{105}.
\bibitem{witten} E. Witten, hep-th/9802150, Adv. Theor. Math. Phys. 2 (1998) 253.
\bibitem{henningson} M. Henningson and K. Skenderis, hep-th/9806087, J. High Energy Phys. 07 (1998) 023.
\bibitem{sn} J. Navarro-Salas and P. Navarro, hep-th/9807019, \plb{439}{1998}{262}
\bibitem{polyakov} A. M. Polyakov, {\it Gauge Fields and Strings}, Harwood
Academic Publishers (1987).
\bibitem{af} G. E. Arutyunov and S. A. Frolov, hep-th/9806216.
\bibitem{henneaux} O. Coussaert and M. Henneaux, hep-th/9310194, \prl{72}{1994}{183}.
\bibitem{wald} R. M. Wald, {\it General Relativity}, The Univ. of Chicago Press.
 
\end{thebibliography}
\end{document}